\documentclass[fleqn,usenatbib]{mnras}
\usepackage{newtxtext,newtxmath}
\usepackage{setspace}
\usepackage[T1]{fontenc}
\usepackage{ae,aecompl}

\newcommand{\Msun}{\mbox{M$_{\odot}$}}
\newcommand{\Rsun}{\mbox{R$_{\odot}$}}

\newcommand{\gppr}{\stackrel{>}{\scriptstyle \sim}}
\newcommand{\gappr}{\raisebox{-0.4ex}{$\gppr$}}
\newcommand{\lppr}{\stackrel{<}{\scriptstyle \sim}}
\newcommand{\lappr}{\raisebox{-0.4ex}{$\lppr$}}

\usepackage[usenames,dvipsnames]{xcolor}
\usepackage{color,colortbl}
\usepackage{soul}
\usepackage{graphicx}	% Including figure files
\usepackage{amsmath}	% Advanced maths commandsl
\defcitealias{parsons16}{Paper~I}
\defcitealias{Rebassa-Mansergas17}{Paper~II}
\defcitealias{lagosetal20-1}{Paper~III}
\title[Three close binaries with early type secondary]{The White Dwarf Binary Pathways Survey IV: Three close white dwarf binaries with G-type secondary stars}

\author[M.-S. Hernandez et al.]{
M.S. Hernandez$^{1}$,\thanks{E-mail: mercedes.hernandez@postgrado.uv.cl}
M.R. Schreiber$^{2,3}$,
S.G. Parsons$^{4}$,
B.T. Gänsicke$^{5,6}$,
F. Lagos$^{1,3,7}$,\newauthor
R. Raddi$^{8}$,
O. Toloza$^{5,6}$,
G. Tovmassian$^{9}$,
M. Zorotovic$^{1}$,
P. Irawati$^{10}$,
E. Past{\'e}n$^{1,3}$,\newauthor
A. Rebassa-Mansergas$^{8,11}$,
J.J. Ren$^{12}$,
P. Rittipruk$^{10}$,
C. Tappert$^{1}$
\\
$^{1}$Instituto de F{\'i}sica y Astronom{\'i}a de la Universidad de Valpara{\'i}so, Av. Gran Breta\~na 1111, Valpara{\'i}so, Chile.\\
$^{2}$Departamento de F{\'i}sica, Universidad T\'ecnica Federico Santa Mar\'ia, Av. España 1680, Valpara{\'i}so, Chile.\\
$^{3}$Millennium Nucleus for Planet Formation, NPF, Valpara{\'i}so, Chile\\
$^{4}$Department of Physics and Astronomy, University of Sheffield, Sheffield S3 7RH, UK.\\
$^{5}$University of Warwick, Department of Physics, Gibbet Hill Road, Coventry, CV4 7AL, United Kingdom.\\
$^{6}$Centre for Exoplanets and Habitability, University of Warwick, Coventry CV4 7AL, UK.\\
$^{7}$European Southern Observatory (ESO), Alonso de C{\'o}rdova 3107, Vitacura, Santiago, Chile\\
$^{8}$Departament de F{\'i}sica, Universitat Polit{\`e}cnica de Catalunya, c/Esteve Terrades 5, E-08860 Castelldefels, Spain.\\
$^{9}$Instituto de Astronom{\'i}a, Universidad Nacional Aut{\'o}noma de M{\'e}xico, Apartado Postal 877, Ensenada, Baja California, 22800, M{\'e}xico.\\
$^{10}$National Astronomical Research Institute of Thailand, Sirindhorn AstroPark, Donkaew, Mae Rim, Chiang Mai 50180, Thailand.\\
$^{11}$Institute for Space Studies of Catalonia, c/Gran Capit{\`a} 2–4, Edif. Nexus 201, 08034 Barcelona, Spain.\\
$^{12}$Key Laboratory of Space Astronomy and Technology, National Astronomical Observatories, Chinese Academy of Sciences, Beijing 100101, P. R. China\\
}

\date{Accepted 2020 December 04. Received 2020 December 03; in original form 2020 July 17}

\pubyear{2019}

\begin{document}
\label{firstpage}
\pagerange{\pageref{firstpage}--\pageref{lastpage}}
\maketitle

% Abstract of the paper
\begin{abstract}
Constraints from surveys of post common envelope binaries (PCEBs) consisting of a white dwarf plus an M-dwarf companion have led to significant progress in our understanding of the formation of close white dwarf binary stars with low-mass companions. The white dwarf binary pathways project aims at extending these previous surveys to larger secondary masses, i.e. secondary stars of spectral type AFGK. Here we present the discovery and observational characterization of three PCEBs
with G-type secondary stars and orbital periods between $1.2$ and $2.5$\,days.
Using our own tools as well as MESA we estimate the evolutionary history of the binary stars and predict their future. We find a large range of possible evolutionary histories for all three systems and identify no indications for differences in common envelope evolution compared to PCEBs with lower mass secondary stars.  
Despite their similarities in orbital period and secondary spectral type, we estimate that the future of the three systems are very different: TYC\,4962-1205-1 is a progenitor of a cataclysmic variable system with an evolved 
donor star, 
TYC\,4700-815-1 will run into dynamically unstable mass transfer that will cause the two stars to merge, and
TYC\,1380-957-1 may appear as super soft source before becoming a rather typical cataclysmic variable star.
   
\end{abstract}

% Select between one and six entries from the list of approved keywords.
% Don't make up new ones.
\begin{keywords}
binaries: close -- stars: white dwarfs --
stars: evolution --
techniques: radial velocities
\end{keywords}

%%%%%%%%%%%%%%%%%%%%%%%%%%%%%%%%%%%%%%%%%%%%%%%%%%

%%%%%%%%%%%%%%%%% BODY OF PAPER %%%%%%%%%%%%%%%%%%
\section{Introduction}

Close white dwarf binary stars produce some of the most important 
incidents in modern astronomy. In particular, thermonuclear explosions such as 
Type Ia Supernovae (SN Ia) are one possible outcome of close white dwarf binary star evolution.  
%Since decades mainly two formation channels for SN\,Ia are discussed, 
For decades two main SN\,Ia formation channels were considered,
the single \citep{webbink84-1} and double \citep{iben+tutukov84-1} degenerate channels. Both channels lead to thermonuclear explosions of white dwarfs close to the Chandrasekhar mass limit. 

In recent years it has become evident that reaching a mass close to the Chandrasekhar limit might not be a strictly necessary criterion for SN\,Ia explosions, enlarging the possible evolutionary scenarios for single or double degenerate SN\,Ia detonations  \citep{finketal07-1,simetal10-1,kromeretal10-1,brooksetal16-1,guillochonetal10-1,vankerkwijketal10-1,pakmoretal13-1}. Furthermore, although classical SN\,Ia lying on the Phillips relation \citep{nugentetal95-1} still make up the majority of observed thermonuclear SNe explosions, a variety of peculiar SNe related to the thermonuclear explosion of a white dwarf have been discovered recently. Among others, these atypical thermonuclear SNe include 
explosions with low ejecta velocities (e.g. SN\,Iax), possibly produced by accretion of a white dwarf from a He-star
that may potentially leave a burned white dwarf remnant \citep{foleyetal13-1},  
%\textbf{no me queda claro esta relacion, creo que va un punto y son cosas separadas, no? Si, esta claro!!} 
and calcium rich SNe which might be long delay-time thermonuclear explosions of a white dwarf in a binary system that was dynamically ejected from its host galaxy \citep{foley15-1}.  
%{\textbf{Furthermore,.... (discuss the briefly the different flavours of thermonueclear supernovae)}}
The growing variety of potential SN\,Ia progenitor systems 
and the increasing number of different flavours of thermonuclear supernovae \citep[see][for a review]{jhaetal19-1}, 
all resulting from close binaries that contain
at least one white dwarf driven close to the Chandrasekhar mass limit, imply that a detailed understanding of white dwarf binary star evolution is required if we want to know 
what are the main progenitor systems of thermonuclear SN and how frequently a given SN\,Ia progenitor configuration is produced.  

The standard hypothesis for close white dwarf binary formation is that the initial main sequence binaries, consisting of two stars with masses exceeding $1\,\Msun$, 
evolve through a common envelope phase
\citep[e.g.][]{webbink84-1,zorotovicetal10-1} generated by 
the initially more massive star filling its Roche-lobe as a giant star. 
The close binaries emerging from common envelope evolution 
consist of a white dwarf and a main sequence companion star with an orbital period usually shorter than a few days.     
%The precise orbital separation of a recently formed post common %envelope binary (PCEB) largely determines 
The future evolution of these post common envelope binaries (PCEBs) depends essentially on their orbital period but also on 
the evolutionary status of the secondary star and the mass ratio. Broadly speaking, for periods longer than $1$-$2$~days, a second phase of mass transfer will likely be initiated when the secondary has evolved into a giant star which will lead in most cases to a second common envelope phase and potentially to a double degenerate binary or a merger \citep[e.g.][]{webbink84-1,olligetal15-1}. For shorter periods, angular momentum loss can drive the system into mass transfer when the secondary star is still on the main sequence which, depending on the mass ratio, can generate thermal time scale mass transfer and stable nuclear burning on the surface of the white dwarf \citep{shen+bildsten07-1}. These systems are called Super Soft X-ray Sources (SSS) and are considered possible single degenerate SN\,Ia progenitors as the white dwarf can grow in mass \citep[e.g.][]{distefano10-1}. 

Despite the general understanding of close white dwarf binary formation outlined above, the predictions of present day binary population models for close white dwarf binaries are very uncertain.  
Usually the outcome of common envelope evolution is determined 
using a simple parameterized energy conservation equation 
and, unfortunately, the key parameter, the common envelope efficiency, is so far only well constrained for low-mass ($\lappr0.5~\Msun$) main sequence secondary stars 
\citep{zorotovicetal10-1,Nebot11}. These low mass PCEBs are important for population studies of cataclysmic variables (CVs) 
but most likely not relevant for studying evolutionary channels towards SN\,Ia explosions because, as a consequence of nova eruptions, the white dwarfs in CVs do not significantly grow in mass \citep{yaronetal05-1,wijnenetal15-1,schreiberetal16-1}. 
%the total mass of these systems hardly reaches 
%the Chandrasekhar limit. 
Providing observational constraints on the formation and evolution of close white dwarf binaries with more massive secondary stars is therefore of utmost importance for a better understanding of binary pathways that may lead to thermonuclear SN.  

We are conducting a survey of close binaries consisting of main sequence stars of spectral type AFGK plus a white dwarf (WD+AFGK from now on). 
Our target selection is based on optical surveys that are combined with the Galaxy Evolution Explorer DR6+7
photometry \citep[\textit{GALEX};][]{bianchi14}, 
to identify objects with ultraviolet (UV) excess indicative for the presence of a white dwarf. 
Close binaries are then identified and characterized 
through radial velocity measurements. So far, we have identified the first pre-SSS binary
system \citep{parsons15}, confirmed that our target selection is 
efficient \citep[][hereafter paper I]{parsons16}, obtained first measurements of 
the fractions of close binaries among our targets  \citep[][hereafter paper II]{Rebassa-Mansergas17}, 
and discussed in detail the limited contamination of our target 
sample through hierarchical triple systems with a white dwarf component \citep[][hereafter paper III]{lagosetal20-1}.  

Here we present the identification and analysis of three close white dwarf binaries with early type secondary stars. We measured their orbital periods,
estimated the stellar masses based on high resolution spectroscopy and used numerical codes to describe the history and future of the three systems. 
We found that these three close binaries have most likely formed through common envelope evolution and that their evolutionary history can be reconstructed with common envelope parameters similar to those used for PCEBs with M-dwarf companions.
Concerning the future of the three systems we obtained that, despite having relatively similar stellar and binary parameters, their final fates are rather different.

\section{Observations and data analysis}

White dwarf plus AFGK binaries are very difficult to be identified based on optical data alone because the main sequence star completely outshines any contribution from the white dwarf.  Here we present a brief explanation of how we overcame this problem and describe the identification process of the three objects discussed in this work.
%The three objects discussed in this work have been selected from cross matching optical surveys with \textit{GALEX} observations to identify stars with UV excess that indicates the presence of a white dwarf. Spectroscopic follow up observations are then required to find the close binaries among this sample of WD+AFGK stars and to measure their periods. 

\subsection{Target selection}

\begin{figure}
	% To include a figure from a file named example.*
	% Allowable file formats are eps or ps if compiling using latex
	% or pdf, png, jpg if compiling using pdflatex
	\includegraphics[width=1.0\columnwidth]{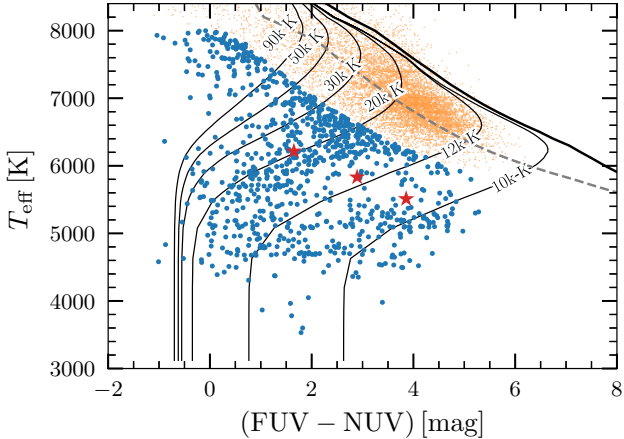}
    \caption{ \textit{GALEX} ${\rm FUV}-{\rm NUV}$ colours vs photometric $T_{\rm eff}$ of $\sim\,13\,000$ TGAS stars (orange density cloud). The WD+AFGK binary candidates are shown as blue circles, while the three binaries studied in this paper are represented by red star symbols. The synthetic colours of {\em PHOENIX} \citep{Husser13} models with $\log{g} = 4.5$ and $Z=0$, or $\log{g}=3.5$ and $Z = -1$, are displayed as solid black and dashed grey curves, respectively. The colours of WD+AFGK companions, computed for a representative set of white dwarf $T_{\rm eff}$, are determined by combining $\log{g} = 8$, \citet{koester10-1} models with {\em PHOENIX} models, which we scaled via the WD mass-radius relation \citep{fontaine01} and the main-sequence masses and radii \citep{choi16}, respectively. }
    \label{fig:galex}
\end{figure}

 Whereas we relied in  \citetalias{parsons16} and \citetalias{ Rebassa-Mansergas17} on the selection of WD+AFGK binary candidates from the RAdial Velocity Experiment \citep[RAVE;][]{kordopatisetal13-1} and The Large Sky Area Multi-Object Fibre Spectroscopic Telescope  \citep[LAMOST;][]{dengetal12-1} 
 databases, in the present work, we have taken advantage of the {\em Tycho-Gaia} Astrometric Solution 
 \citep[{\em TGAS};][]{Lindegren16} for the identification of suitable candidates in a magnitude limited sample.

For the purpose of characterising the excess emission with respect to normal UV colours of AFGK stars, we cross-matched the TGAS database with DR6+7 \textit{GALEX} photometry, 
identifying $\sim\,13\,000$ reliable matches with ${\rm FUV} \leq 19$ and TGAS parallax uncertainty $\sigma_\varpi/\varpi \leq 0.15$.
We complemented them with optical and infrared photometry from The AAVSO Photometric All-Sky Survey DR9 \citep[APASS;][]{Henden15}, Two Micron All-Sky Survey \citep[{\em 2MASS};][]{Skrutskie06}, and Wide-Field Infrared Survey Explorer   \citep[{\em WISE};][]{Wright10}. We determined photometric effective temperatures, $T_{\rm eff}$, by fitting the optical-to-infrared spectral energy distributions (SED) with a grid of synthetic spectra \citep[PHOENIX;][]{Husser13}. The result of the SED fitting delivered $\approx 1000$ WD+AFGK binary candidates, displaying moderate to significant UV-excess with respect to normal colours of main-sequence and giant stars of below-Solar metallicity ($Z = -1$).
We later revised the target selection by also including parallaxes and broad-band photometry 
from the second data release of {\em Gaia} \citep[DR2;][]{gaiaetal18-1}, bringing the target list down to $\approx 800$ candidates.
The identification of suitable candidates is illustrated in Fig.\,\ref{fig:galex}. A more detailed description of the photometric classification of the candidates and first spectroscopic follow-up observations of the sample, are presented in \citet{ren20}.

%We performed follow-up observations searching for radial velocity variations in a large number of systems. 
Here we present follow-up observations and a detailed analysis of three systems with short orbital periods: TYC\,4700-815-1, TYC\,1380-957-1 and TYC\,4962-1205-1. The \textit{GALEX} magnitudes  with APASS, 2MASS and WISE fluxes of these three systems are listed in Table \ref{tab:galex}.
Hereafter, we will refer to the three systems as TYC\,4700, TYC\,1380 and TYC\,4962 respectively. 

\begin{table}
	%\centering
	\caption{\textit{GALEX} magnitudes with APASS, 2MASS and WISE fluxes of the three objects discussed in this work}
	\label{tab:galex}
	\resizebox{\columnwidth}{!}{\begin{tabular}{lcccc}
	%\begin{tabular}{lcccc} % four columns, alignment for each
		\hline
Filter& $\lambda$ [\AA] & TYC\,4700-815-1& TYC\,1380-957-1& TYC\,4962-1205-1\\
\hline
FUV & 1516 &14.91$\pm 0.01$ & 18.31$\pm 0.08$ &18.71$\pm 0.07$ \\
NUV & 2267 &13.26 $\pm 0.00$ & 15.45$\pm 
0.01$ &14.51$\pm 0.01$ \\
\hline
V &5\,394.3 & 9.36(0.06)$\times10^{-13}$ & 2.06(0.09)$\times10^{-13}$ & 9.69(0.04)$\times10^{-13}$ \\
J & 12\,350 & 2.05(0.04)$\times10^{-13}$  & 5.14(0.01)$\times10^{-14}$ & 2.80(0.05)$\times10^{-13}$\\
H & 16\,620 & 9.25(0.04)$\times10^{-14}$  & 2.36(0.67)$\times10^{-14}$  & 1.52(0.06)$\times10^{-13}$ \\
Ks & 21\,590 &  3.67(0.71)$\times10^{-14}$ & 1.00(0.20)$\times10^{-14}$ & 6.12(0.02)$\times10^{-14}$\\
W1 & 33\,526 & 7.47(0.16)$\times10^{-15}$  & 1.93(0.004)$\times10^{-15}$ & 1.25(0.36)$\times10^{-14}$ \\
W2 & 46\,028 & 2.14(0.039)$\times10^{-15}$  &5.50(1.01)$\times10^{-16}$ &  3.66(0.006)$\times10^{-15}$ \\
	\hline
	\end{tabular}}
	V, J, H, Ks, W1, W2 fluxes are in $erg~cm^{-2}s^{-1}$\AA$^{-1}$ \\
\end{table}

\subsection{High resolution spectroscopy} 

The high-resolution spectroscopic observations were performed using five different instruments installed at different telescopes. We used the Ultraviolet and Visual Echelle Spectrograph \citep[UVES,][]{Dekker00} on the Very Large Telescope in Cerro Paranal and the Fiber-fed Extended Range Optical Spectrograph \citep[FEROS,][]{Kaufer99} at the $2.2$~m telescope of the Max Planck Gesellschaft (MPG) at La Silla Observatory, both in Chile. We furthermore obtained data with the High Resolution Spectrograph \citep[HRS,][]{Jiang02} at Xinglong $2.16$~m telescope (XL216) located in China, the Middle-Resolution fiber-fed Echelle Spectrograph (MRES) at the {\bf $2.4$\,m} Thai National Telescope (TNT)  in Thailand and the Echelle SPectrograph from REosc for the Sierra San pedro martir Observatoy  \citep[ESPRESSO,][]{levin95} at $2.12$~m telescope of the Observatorio Astron\'omico Nacional at San Pedro M\'artir (OAN-SPM)\footnote{http://www.astrossp.unam.mx}, M\'exico. Nearly all spectra were obtained during gray nights and relatively poor weather conditions. However, all with sufficient signal to noise ratios. % of at least $30$ were obtained in all cases.}

The cross-dispersed echelle spectrograph UVES has a spectral resolution of $58\,000$ for a $0.7$~arcsec slit. The two-arms cover the wavelength range of $3000$--$5000$~\AA\,  (Blue) and $4200$--$11\,000$~\AA\, (Red), centered at $3900$ and $5640$~\AA\, respectively. Standard reduction was performed making use of the specialized pipeline EsoReflex workflows \citep{Freudling13}.  

FEROS on the 2.2~m Telescope has a resolution of $R \approx48\,000$ and covers the wavelength range from  $\sim$ $3500$--$9200$~\AA. Data obtained with the FEROS spectrograph were extracted and analysed with an automated pipeline developed to process spectra coming from different instruments in an homogeneous and robust manner, the CERES code \citep[][]{jordan14,Brahm17}. After performing typical image reductions, spectra were optimally extracted following \citet{Marsh89}. The instrumental drift in wavelength through the night was corrected with a secondary fiber observing a Th-Ar lamp.

The HRS Spectrograph at Xinglong 2.16~m telescope provides echelle spectra of a $49\,800$ resolving power for a fixed slit width of $0.19$\,mm and covers a wavelength range of $\sim$\,$3650$--$10\,000$~\AA. \mbox{ Th-Ar}~arc spectra were taken at the beginning and end of each night.
Additional follow-up spectroscopy with  
MRES at the TNT was obtained using a slit width of $1.4$~arcsec. MRES provides spectra of a resolving power of $15\,000$ covering the $3900$--$8800$~\AA\,  wavelength range. Arc spectra were taken at the beginning of each night and we used sky lines to account for the flexure of the spectrograph.

The ESPRESSO Spectrograph of OAN-SPM provides spectra covering $3500$--$7105$~\AA\, with a spectral resolving power of $ R\approx 18\,000$. A Th-Ar lamp before each exposure was used for wavelength calibration. 
The MRES, HRS and ESPRESSO spectra were reduced using the \textit{echelle} package in IRAF\footnote{IRAF is distributed by the National Optical Astronomy Observatories, which are operated by the Association of Universities for Research in Astronomy, Inc., under cooperative agreement with the National Science Foundation.}. Standard procedures, including bias subtraction, cosmic-ray removal, and wavelength calibration were carried out using the corresponding tasks in IRAF.

\subsection{Radial Velocity Measurements}
\label{sec:maths} % used for referring to this section from elsewhere

Radial velocities (RVs) from UVES and ESPRESSO spectra were computed using the cross-correlation technique against a binary mask %representative of the
corresponding to a G-type star. 
The uncertainties in radial velocity were computed using scaling relations \citep[][]{jordan14} with the signal-to-noise ratio and width of the cross-correlation peak, which were calibrated with Monte Carlo simulations.
FEROS spectra were analyzed with the CERES code which also calculates radial velocities using cross-correlation
\citep[for more details, see][]{jordan14,Brahm17}. 
RVs from the spectra obtained with HRS and MRES telescopes were obtained by fitting the normalised Ca{\sc ii} absorption triplet (at $8498.02, 8542.09$, and $8662.14$~\AA) and Na{\sc i} respectively, with a combination of a second order polynomial and a triple-Gaussian profile of fixed separations, as described in \citetalias{Rebassa-Mansergas17}. Only when the Ca{\sc ii} absorption triplet was too noisy to get a reliable RV, the Na{\sc i} doublet at $\sim 5889.95$ and $5895.92$~\AA\,was used. 
In the latter case we used a second order polynomial and a double-Gaussian profile of fixed separation. The RV uncertainty is obtained by summing the fitted error and a systematic error of $0.5$~km~s$^{-1}$ in quadrature (more details in \citet{ren20}). 

All RVs obtained for the three objects are provided in the appendix in the 
tables \ref{tab:TYC4700}, \ref{tab:TYC1380}, \ref{tab:TYC4962}.

\section{Stellar and binary parameters}

\subsection{Characterizing the secondary star} 
\label{sec:secondary}

\begin{figure}
    \centering
    \includegraphics[width=1\columnwidth]{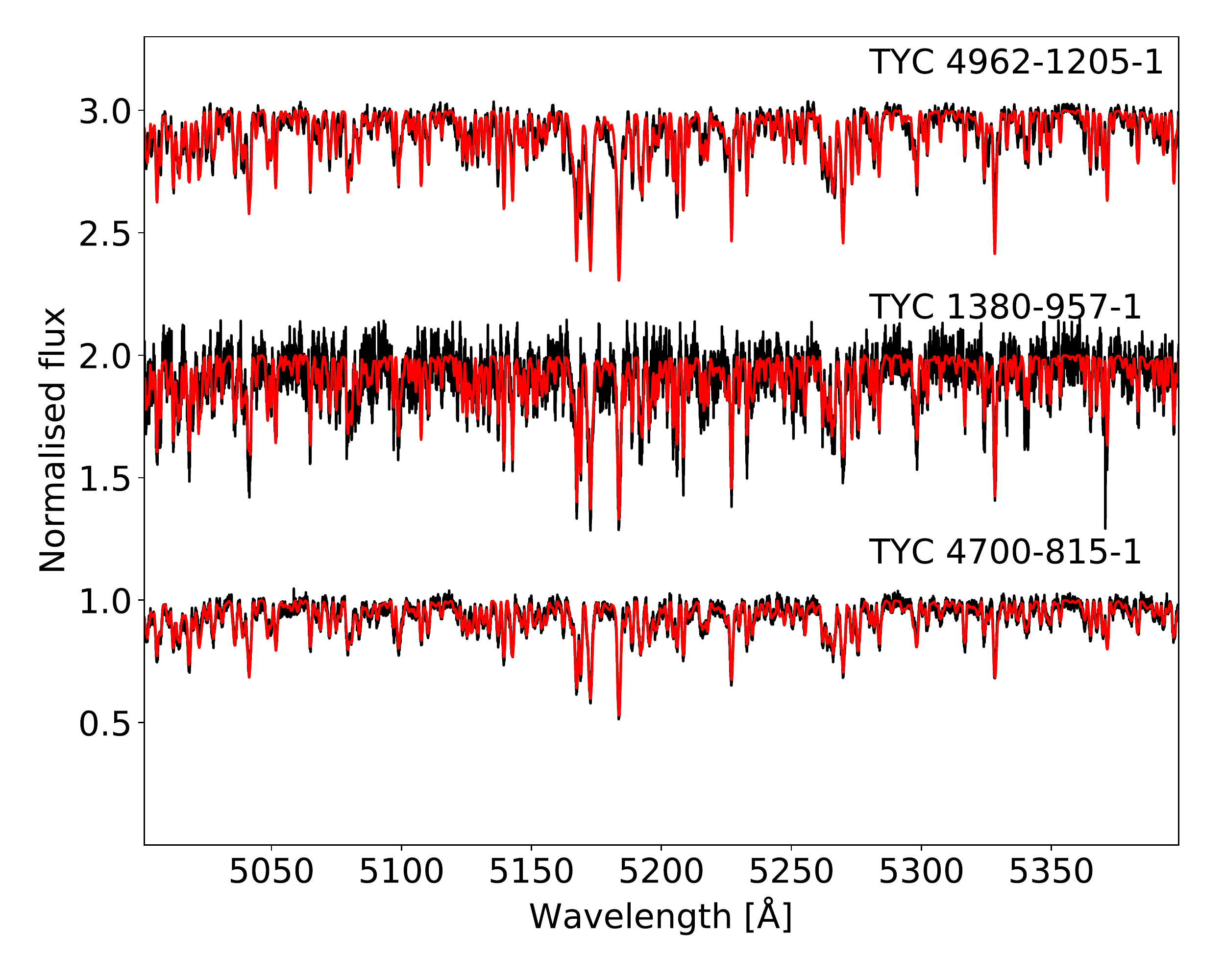}
    \caption{A small range of the observed high resolution spectra (black) from UVES for TYC\,4700 and FEROS for TYC\,1380 and TYC\,4962. Spectral fits to the three G type stars using synthetic spectra within iSpec (red). The best-fit results were combined with SED fitting and {\it Gaia} DR2 parallaxes to obtain our final parameter estimates. }
    \label{fig:specfit}
\end{figure}

We determined the parameters of the main-sequence stars in our binaries by fitting our high-resolution spectra with model spectra and combining the obtained results with a fit to their spectral energy distributions (SEDs) scaled to {\it Gaia} DR2 parallaxes.
This way, the temperature and surface gravities of the secondary stars are constrained from the spectral fit. The radius of the secondary star is estimated by scaling a theoretical model %to the Gaia parallax 
to match the SEDs fluxes at the distance implied by the {\it{Gaia}} parallaxes. 

 For each target we selected the spectra with the highest signal-to-noise ratio and normalised the  continuum  of the spectra. Based on the initial parameter estimates of the stars from the CERES pipeline, synthetic spectra  were generated with iSpec
\citep{Blanco2014}. We then used the synthetic spectral fitting technique within iSpec to measure the effective temperatures, surface gravities, metallicities [Fe/H] and rotational broadenings for our targets. The limb darkening coefficient was kept fixed at a value of $0.6$ and the spectral resolution was fixed at the measured value for each observation based on fits to sky lines. A small section of the spectra and the best fit models are shown in Fig.\,\ref{fig:specfit}.

\begin{figure*}
    \centering
    	\includegraphics[width=1.8\columnwidth]{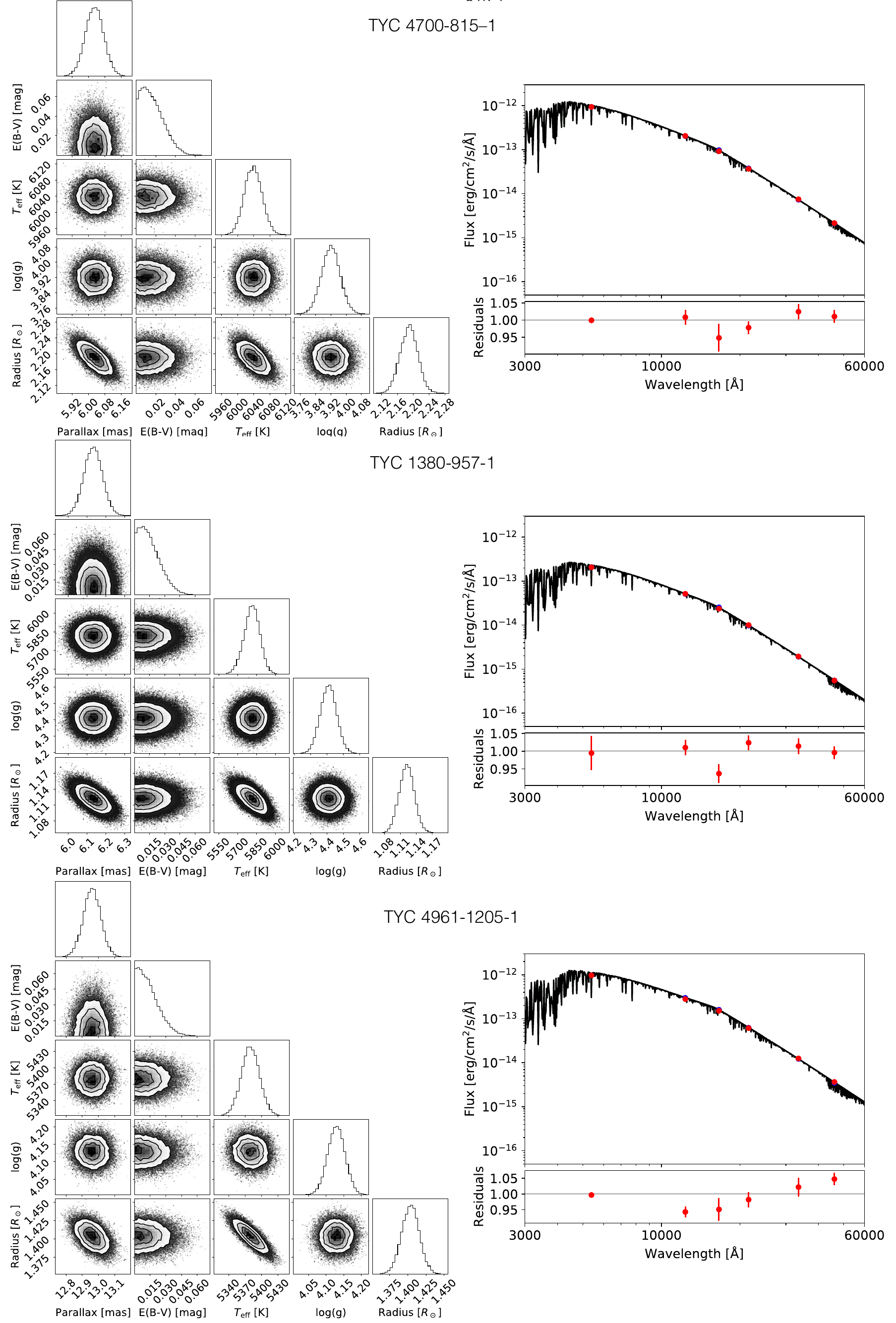}
    	%1.8  0.9
    \caption{
    Spectral energy distributions created from AAVSO Photometric All Sky Survey, $J$, $H$ and $K_s$ band data from {\rm{2MASS}}, and $W1$ and $W2$ band data from WISE were fitted with BT-NextGen theoretical spectra. We used as input parameters the previously estimated effective temperature, surface gravity and radius of the star and the parallax and reddening of the system. The Markov chain Monte Carlo method was used to derive posterior distributions from which we derived our final parameters and the corresponding uncertainties. 
    }
    \label{fig:sed}
\end{figure*}

 SEDs of our targets including $V$ band data from the {\em APASS} \citep[DR9,][]{Henden15}, $J$, $H$ and $K_s$ band data from {\rm{2MASS}} \citep{Cutri03} and $W1$ and $W2$ band data from 
{\em WISE}  \citep{Cutri12}. We fitted the SEDs with BT-NextGen theoretical spectra \citep{Allard12} using a custom Python code. The input parameters of the code are the effective temperature, surface gravity and radius of the star and the parallax and reddening of the system. The theoretical spectra were scaled by a factor $(R/D)^2$, where $R$ is the radius of the star derived from the SED we create of the secondary star, $D$ is the distance. The distributions of our model parameters were found using the Markov chain Monte Carlo (MCMC) method \citep{Press07} implemented using the python package {\sc emcee} \citep{Foreman13}, where the likelihood of accepting a model was based on the $\chi^2$ of the fit and additional prior probabilities on the effective temperature and surface gravity from the iSpec fit, the parallax from {\it Gaia} DR2 \citep{Gaia18} and the reddening from 3D interstellar dust maps \citep{Lallement19}. A short initial MCMC chain was used to determine the approximate parameter values, which were then used as the starting values in a longer "production" chain, to determine the final values and their uncertainties. The production chain used $50$ walkers, each with $10\,000$ points. The first $1000$ points were classified as "burn-in" and were removed from the final results. Stellar masses were then computed using the fitted surface gravities and radii. The best fit values and their uncertainties are listed in Table \ref{tab:stellarP} and plotted in Fig.\,\ref{fig:sed}, along with the parameter distributions.

\subsection{Orbital periods}

\begin{figure}
    \centering
	\includegraphics[width=1\columnwidth]{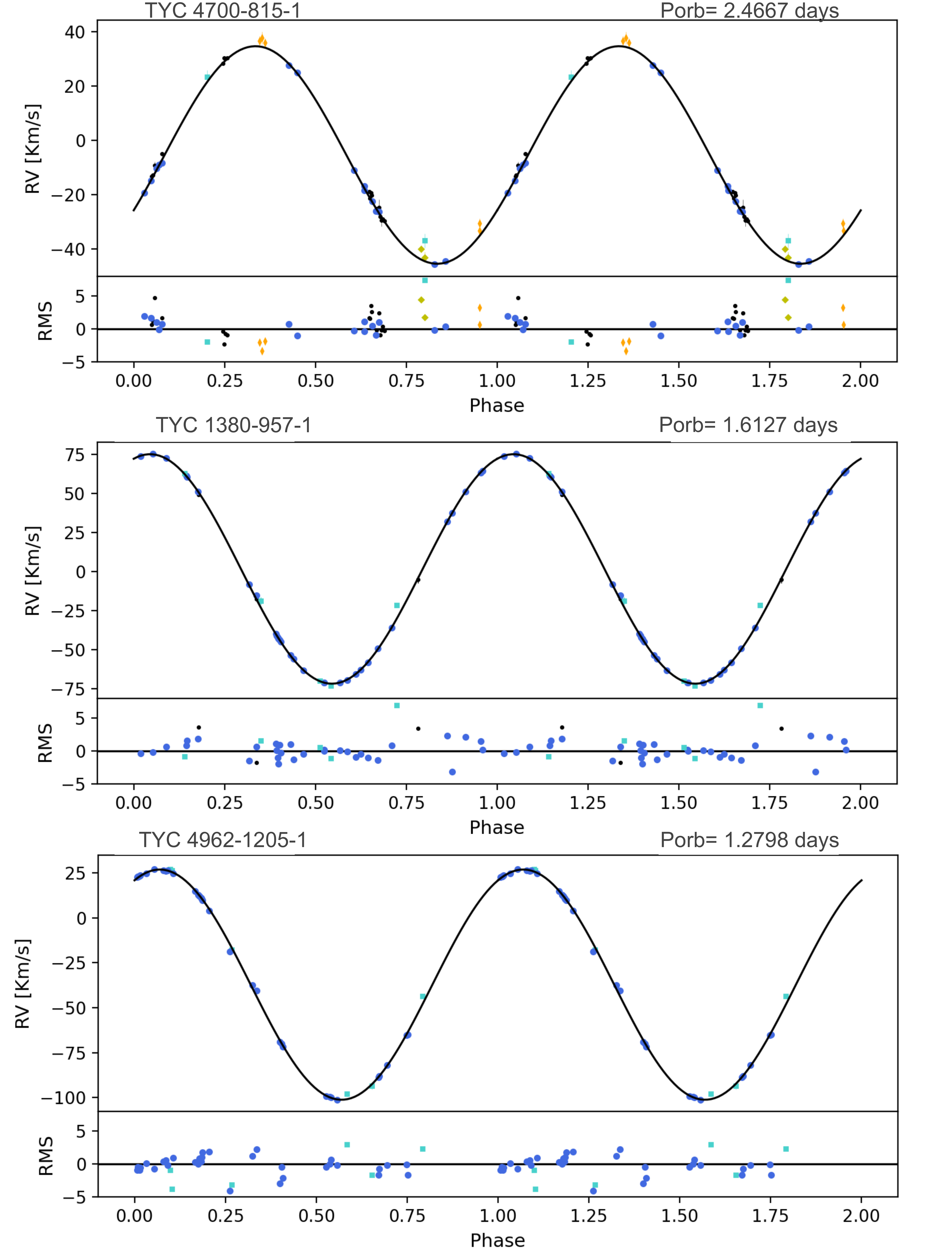}
    \caption{Phase-folded radial velocity curve for the main-sequence stars with the corresponding residuals to the best fit. The colors indicate which instrument was used to obtain the measurement: FEROS (blue big dots), ESPRESSO (cyan squares), HRS (black small dots), MRES (orange thin diamonds), and UVES (green diamonds).}
    \label{fig:RVC}
\end{figure}

\begin{figure}
    \centering
    \includegraphics[width=0.9\columnwidth]{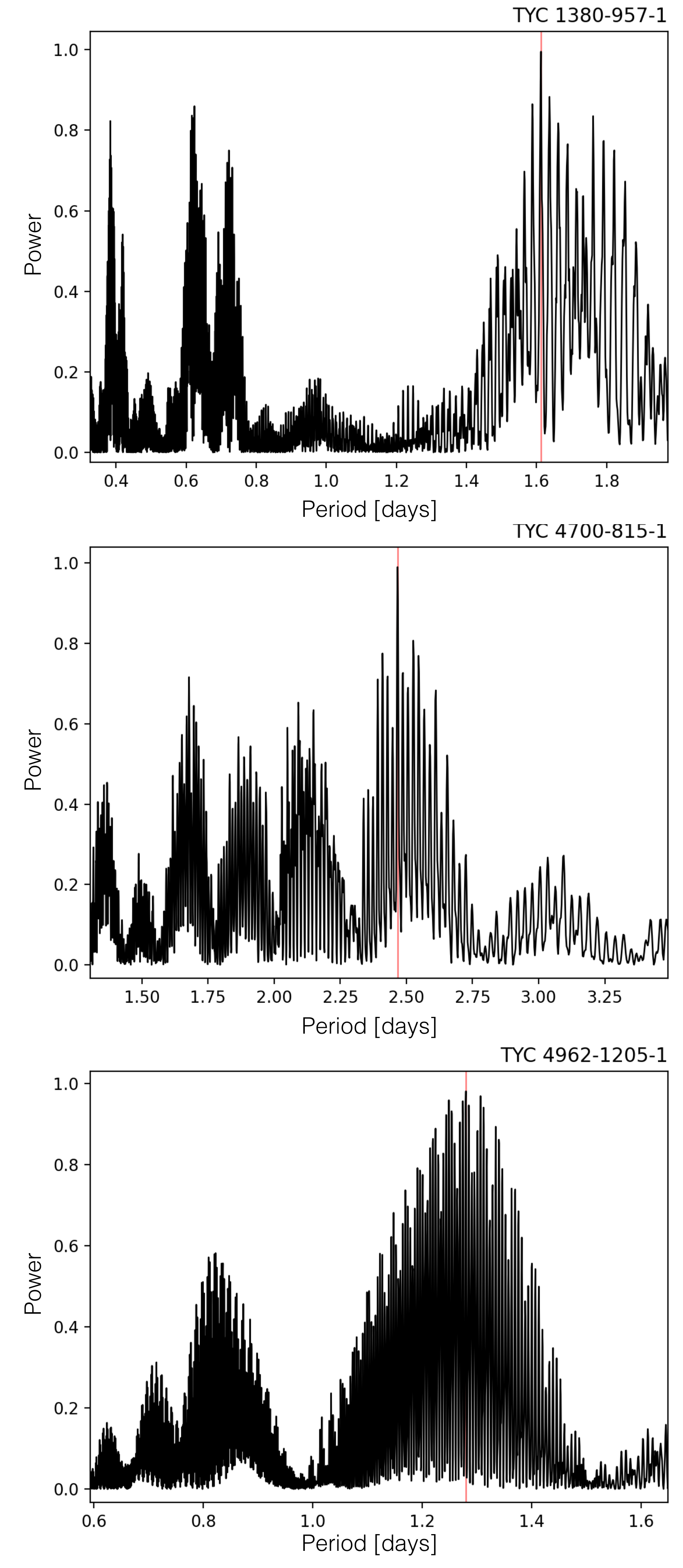}  %width=1.0
    \caption{Lomb-Scargle periodograms of TYC\,1380 (top panel), TYC\,4700 (middle) and TYC\,4962 (bottom). The red lines highlight the highest peak which corresponds to the orbital period of the systems.} 
    \label{fig:periodogram}
\end{figure}

The most critical parameter constraining the past and future of a given PCEB is its orbital period. 
We used the least-squares spectral analysis method based on the classical Lomb-Scargle periodogram \citep{Lomb76,Scargle82} that was implemented in the astroML python library \citep{VanderPlas12} and {\em{The Joker}} \citep{Price-Whelan17, Price-Whelan20}, a specialized Monte Carlo sampler created to find converged posterior samplings for Keplerian orbital parameters.
{\em{The Joker}} is especially useful for non-uniform data and allows to also identify eccentric orbits. As we have previously identified systems with eccentric orbits that most likely represent hierarchical triples with white dwarf components (\citetalias{lagosetal20-1}), double checking circular orbits derived from Lomb-Scargle periodograms with {\em{The Joker}} was required. 
{\em{The Joker}} found the period corresponding to the highest peak in the peridograms shown in  
Fig. \ref{fig:periodogram} and the eccentricity in all three cases to be insignificant. The highest peaks in the periodograms shown in Fig. \ref{fig:periodogram} also clearly provided the best fit to the data. All other aliases with similar power in the periodogram disagreed drastically with some radial velocity measurements and the $\chi^2$ values resulting from sine fits using the periods corresponding to the second highest peaks in the periodograms are at least a factor $33$ higher in all cases and can therefore be excluded. 
The measured orbital periods for TYC\,4700, TYC\,1300 and TYC\,4962 are $2.4667$, $1.6127$ and $1.2798$~days respectively. The final sine fits to our radial velocity data are shown in Fig.~\ref{fig:RVC}. 
Radial velocities from different telescopes are plotted in different colors.

\subsection{The white dwarf mass} 

The masses of white dwarfs in binaries are notoriously difficult to measure and usually UV spectra 
e.g taken with the Hubble Space Telescope (\textit{HST}) are required to get reliable estimates (e.g. \citetalias{parsons16}). %\citep[e.g.][]{parsons16}. 
For the three systems presented here,  \textit{HST} spectra are not available and the presence of a white dwarf is only indicated by excess UV emission detected with \textit{GALEX}. However, based on the parameters estimated for the secondary star we can derive rough estimates for the white dwarf mass. 
A lower limit for the white dwarf mass can be obtained based on the measured secondary mass and orbital period and assuming the maximum inclination of 90 degrees. Assuming furthermore that the orbit of the secondary is synchronised, a reasonable assumption for the short orbital periods we measured, 
we can estimate a range of possible white dwarf masses based on the $v\sin(i)$ and radius measured for the secondary star using the binary mass function. 

Both estimates, the minimum white dwarf mass and the one based on $v\sin(i)$ and synchronised rotation are shown in Table\,\ref{tab:stellarP}. We note that especially the latter estimate must be taken with a grain of salt, as 
the secondary star parameters were obtained by fitting single main sequence star model spectra to observed spectra of 
secondary stars in close binaries. This might imply that our $v\sin(i)$ measurements are affected by systematic errors that most likely exceed the purely statistical errors we provide in Table\,\ref{tab:stellarP}.

\begin{table}
	%\centering
	\caption{Stellar Parameters}
	\label{tab:stellarP}
	\resizebox{\columnwidth}{!}{\begin{tabular}{lccc} % four columns, alignment for each
		\hline
	Parameter	 & TYC\,4700-815-1 & TYC\,1380-957-1 &  TYC\,4962-1205-1 \\
		\hline
    Orbital Period [days] &2.4667 $\pm 0.0087$ & 1.6127 $\pm 0.0109$  & 1.2798 $\pm 0.0026$ \\
    RV amplitude [Km~s$^{-1}$]&  40.067 $\pm 0.083$ &73.32  $\pm 0.025$ & 64.057 $\pm 0.014$ \\
    $V_\gamma$ [Km~s$^{-1}$] &  -5.46 $\pm 0.044$ & 1.70 $\pm 0.020$ & -37.19 $\pm 0.012$\\
    $v\sin(i)$ [Km~s$^{-1}$]& 45.3 $\pm 4.86$ & 30.00 $\pm 2.84$ & 37.30 $\pm 3.11$ \\
    Inclination [deg]& 63-90 & 49-71 & 37-47\\
    Distance  [pc]& 165.84 $\pm 1.20$ & 163.07 $\pm 1.29$ & 77.141 $\pm 0.321$  \\
    Sec. $T_{\mathrm{eff}}$ [K] & 6038.7 $\pm 31.7 $& 5815.5 $\pm 64.7$ &5380 $\pm 30$ \\
    Sec. Mass  [M$_\odot$]& 1.454 $\pm 0.171$ & 1.181 $\pm 0.145$ &0.969 $\pm 0.058$ \\
    Sec. Radius [R$_\odot$] &2.190$\pm 0.023 $ & 1.122$\pm 0.014$ & 1.404 $\pm 0.012$ \\
    Sec S Type & G0IV  & G2V  &  G6V   \\
    Sec $\log {g}$   [dex]& 3.920 $\pm 0.049$  & 4.410$\pm 0.050$  &4.129 $\pm0.025$ \\
    Sec. Metallicity  [Z/H]& -0.04 $\pm 0.13$ & -0.03 $\pm 0.07$ & -0.42 $\pm 0.07$ \\
    Sec. Luminosity  [L$_\odot$]& 5.713 $\pm 0.093$ &1.290 $\pm 0.04$ & 1.479 $\pm 0.014$ \\
    Minimum WD Mass  [M$_\odot$]& 0.38 & 0.59 &0.4 \\
    WD Mass$(i)$   [M$_\odot$] & 0.38-0.44 & 0.64-0.85 & 0.59-0.77\\
		\hline
	\end{tabular}}
\end{table}

\section{Discussion} 

We have identified three close binary stars consisting of a G-type star and most likely a white dwarf whose presence is indicated by a significant UV excess, and measured the stellar and binary parameters of the three systems. In what follows we investigate possible implications for white dwarf binary formation and evolution by reconstructing their past and predicting their future. 
Finally, we will discuss how certain we can be that our interpretation that the UV excess comes from a white dwarf is the most likely one. 

\subsection{The evolutionary history}

The main purpose of the white dwarf binary pathway project is to 
progress with our understanding of the formation and evolution of close white dwarf binaries with secondary stars in the mass range $\sim0.5$--$1.5$~\Msun. Previous surveys of close white dwarf main sequence binaries nearly exclusively identified systems with M-dwarf companions. 
These surveys were fundamental to constrain the models for 
common envelope evolution. The classical energy conservation 
equation that relates the energy required to unbind the envelope and the released orbital 
energy can be written as  
\begin{equation}
    E_{\mathrm{bind}} = \alpha_{\mathrm{CE}} \Delta E_{\mathrm{orb}}
    %+ \alpha_{\mathrm{rec}} E_{\mathrm{rec}}
\end{equation}
where $\alpha_{\mathrm{CE}}$ is the common envelope efficiency. 
This equation is usually used to estimate the outcome of common envelope evolution. 
Typically a simple expression for the binding energy is used, i.e. 
\begin{equation} 
E_{\mathrm{bind}}=-\frac{GM_1M_\mathrm{1,e}}{\lambda R_1} 
\label{eq_bind}
\end{equation}
where $M_\mathrm{1}$, $M_\mathrm{1,e}$ and $R_\mathrm{1}$ are the total mass, envelope mass and radius of the primary star at the onset of the common envelope phase, respectively.
%has been used assuming a constant binding energy parameter $\lambda$. 
The binding energy parameter $\lambda$ strongly depends on the mass and the radius of the 
white dwarf progenitor when it fills the Roche-lobe and dynamically unstable mass transfer starts.   
A more general form of the binding energy equation takes into account 
possible contributions from recombination energy ($U_\mathrm{rec}$) in the envelope ejection process and a corresponding 
second common envelope parameter $\alpha_{\mathrm{rec}}$, i.e. 
\begin{equation}\label{eq:Eball}
E_\mathrm{bind}=\int_{M_\mathrm{1,c}}^{M_\mathrm{1}}-\frac{G m}{r(m)}dm + \alpha_{\mathrm{rec}}\int_{M_\mathrm{1,c}}^{M_\mathrm{1}}U_\mathrm{rec}(m). 
\end{equation}

Combining the above equations with stellar evolution tracks, it is possible to estimate the progenitor system parameters 
for a given PCEB and a given common envelope efficiency. 
This technique has been pioneered by \citet{nelmeans+tout05}.  
For low mass secondary stars, 
\citet{zorotovicetal10-1} showed that all known systems can be reconstructed with $\alpha_{\mathrm{CE}}$\,=\,$0.2$--$0.3$. 
Their conclusion that, at least for low-mass secondary stars,
the outcome of common envelope evolution can be reproduced by assuming a low value of $\alpha_{\mathrm{CE}}$ has later been confirmed by various 
comparisons of the predictions of 
binary population synthesis with the observed sample  \citep[e.g.][]{Camacho14,Toonen13,zorotovicetal14-1,zorotovicetal11-1}.

One of the key questions the white dwarf binary pathway project shall answer is whether the outcome of common envelope evolution for systems with more massive secondary stars can be described with the same formalism and the same efficiency parameter that is now well established for systems with M-dwarf secondaries.  
Eight systems with secondary stars earlier than M and with orbital periods that clearly imply that mass transfer must have occurred in the progenitor systems, potentially common envelope evolution, have been previously discovered: IK\,Peg, KOI-3278, SLB1, SLB2, SLB3, and KIC\,8145411 with much longer orbital periods of $\sim$ $21$, $88$, $683$, $728$, $419$, 
and $455$\,days respectively \citep{Wonnacott93,Kruse+agol15,Kawahara18,Masuda19} as well as the shorter orbital period systems V471\,Tau \citep{OBrien01} and TYC\,6760-497-1 \citep{parsons15}, both 
with very short periods of just 12 hours. 
Two additional potential members of the class of detached PCEBs with secondary stars earlier than M are   
V1082 Sgr and GPX-TF16E-48 \citep{Tovmassian18, Krushinsky20}. However, for now we exclude these two systems from our 
list of PCEBs as in the first case the mass transfer rate seems unusually 
high for a detached system and as in the second case the secondary star parameters appear to be still somewhat uncertain. 

\begin{figure}
    \centering
    \includegraphics[width=1\columnwidth]{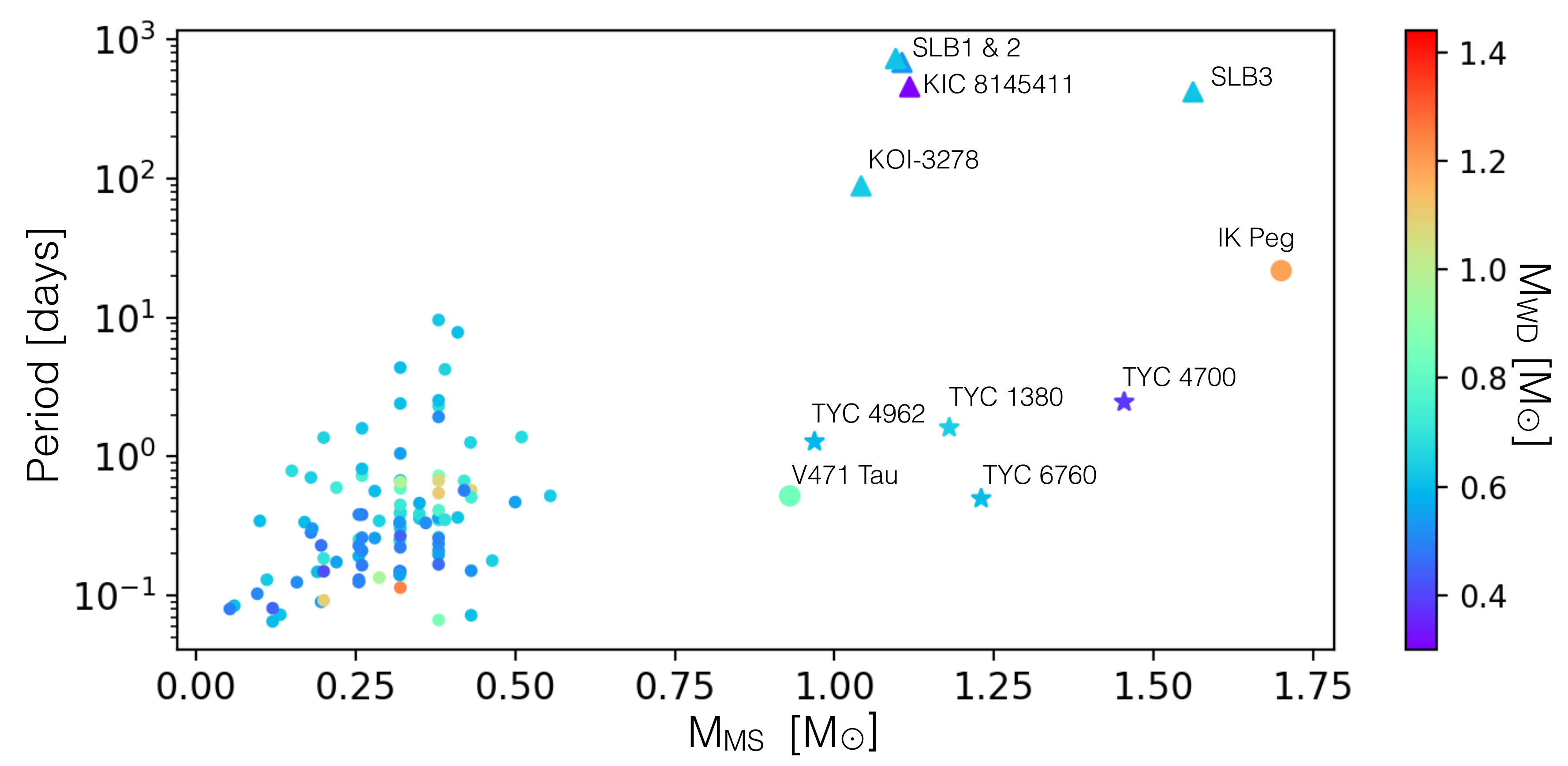}
  \caption{Period distribution of PCEBs including the three systems we present in this work. A large number of PCEBs with M-dwarf companion (small dots) have been characterized and their orbital period distribution peaks at a few hours \citep{Nebot11}. White dwarf binaries with G-type secondary star (the four systems from our pathway survey are represented with stars) cover a much wider range or periods which extends from half a day to several $100$~days. While the short period systems are certainly PCEBS, the origin of the systems with longer periods is less clear. Big circles are previously known PCEBs and triangles are Self-Lensing Binaries.}
    \label{fig:periodD}
\end{figure}

The orbital periods of the currently known potential PCEBs, including the three systems presented here, are shown in Fig.\,\ref{fig:periodD} as a function of secondary mass. For those systems with orbital periods exceeding 20~days, % IK\,Peg and KOI-3278, 
orbital energy as the only energy source during common envelope evolution 
is not sufficient to explain their long orbital periods. 
As shown by \citet{zorotovicetal14-2} such systems require additional energy, most likely recombination energy, to contribute to expelling the envelope. Otherwise, these systems can not be explained in the context of common envelope evolution. 

\begin{table}
	%\centering
	\caption{Binary parameters obtained by reconstructing common envelope evolution where $M_{\mathrm{1,i}}$ and $P_{\mathrm{orb,i}}$ denote the initial primary mass and orbital period. The first row provides the parameters for $\alpha_{\mathrm{CE}}=0.2-0.3$ while in the second row the efficiency was kept as a free parameter.
	}
	\label{tab:rec}
	\resizebox{\columnwidth}{!}{\begin{tabular}{llcccc} % four columns, alignment for each
		\hline
	Object & $\alpha_{\mathrm{CE}}$ & WD Mass [M$_\odot$] &  $M_{\mathrm{1,i}}$ [M$_\odot$] & $P_{\mathrm{orb,i}}$ [days]& Age[Gyr] \\
	%  & & & initial & initial \\ 
	\hline
	TYC 4700 & 0.2--0.3 & 0.40--0.44 & 1.30--1.53 & 203--386 & 2.78--4.81\\
	   & 0.18--1.0 & 0.38--0.44 & 1.29--1.90& 109--395 & \\ 
    TYC 1380 & 0.2--0.3 & 0.64--0.85 & 2.54--3.94 & 706--2222& 0.22--0.77 \\
       & 0.10--1.0 & 0.64--0.85 & 2.36--4.06& 126--2222 & \\
    TYC 4962 & 0.2--0.3& 0.59--0.77 & 2.28--3.36 & 617--1917 & 0.35--1.04\\
       & 0.07--1.0 & 0.59--0.77 & 1.75--3.46 & 103--1917\\
       \hline
	\end{tabular}}
\end{table}

With the three systems presented in this work, our dedicated survey for PCEBs with massive secondary stars has now found four systems with short orbital periods. The evolutionary history of the previously identified PCEB TYC\,6760-497-1 can be reconstructed assuming common envelope evolution without contributions from recombination energy and using the same efficiency that has been shown to work best for PCEBs with M-dwarf secondaries. We applied the algorithm developed by 
\citet{zorotovicetal10-1, zorotovicetal11-1, zorotovicetal14-2}
which combines the stellar evolution code {\it{BSE}} \citep{Hurley02} with the above listed common envelope energy equations
to reconstruct the evolutionary history of the three new
PCEBs with G-type secondary star that we observationally characterized in the previous sections. 

In short, the CE reconstruction works as follows. We calculated a grid of
stellar evolution tracks 
with initial masses up to eight solar
masses (with a step size of $0.01$\Msun). 
As the core mass of the giant star at the onset of CE evolution 
must have been equal to the white dwarf mass in the currently observed post-CE system, 
we searched the grid for giant stars with a core mass equal to the estimated range of possible white dwarf
masses. As the radius and the mass of each of the identified potential
white dwarf progenitor stars is also given by the stellar model, we can determine the orbital period at
the onset of CE evolution for all these candidates using the mass of the companion
and Roche-geometry. Knowing the stellar masses, the orbital period, and the 
envelope mass at the onset of CE evolution as well as the current configuration, allows  
then to determine 
whether sufficient orbital energy was provided to unbind the envelope of the giant and how efficient the use of orbital energy must have been. 

Using this algorithm and leaving the common envelope efficiency as a free parameter we found 
possible progenitor systems for a relatively large range of values for $\alpha_{\mathrm{CE}}$ (see Table\,\ref{tab:rec}). 
While the low white dwarf mass found for TYC\,4700 implies that the progenitor 
filled its Roche-lobe on the first giant branch, for the other two systems two scenarios are possible. Either the progenitor filled its Roche-lobe on the early AGB (low values of $\alpha_{\mathrm{CE}}$) or as a thermally pulsating AGB star ($\alpha_{\mathrm{CE}}\gappr\,0.4$). 

With now five out of eleven %(the four systems from our survey, and excluding V\,471 Tau as it has a K-star companion) 
of all known white dwarf binaries with  secondaries earlier than M having short orbital periods and not requiring extra energy during common envelope evolution, it seems that perhaps a population of PCEBs with such secondaries exists that forms identical to PCEBs with low mass secondaries. A crucial question that remains to be answered is what the origin of the longer orbital period systems is. If it is common envelope evolution as well, then one would need to explain why such long orbital period systems are exclusively found among PCEBs with more massive secondaries. Alternatively, 
at least for some of these longer period systems, it might not have been common envelope evolution that led to their formation. 
However, firm conclusions cannot be drawn from these relatively few  known systems. Therefore, it is of great importance for our understanding of white dwarf binary formation to continue measuring orbital periods of close WD+AFGK binaries. 

\subsection{The future of the three systems} 

\begin{figure}
    \centering
    \includegraphics[width=1\columnwidth]{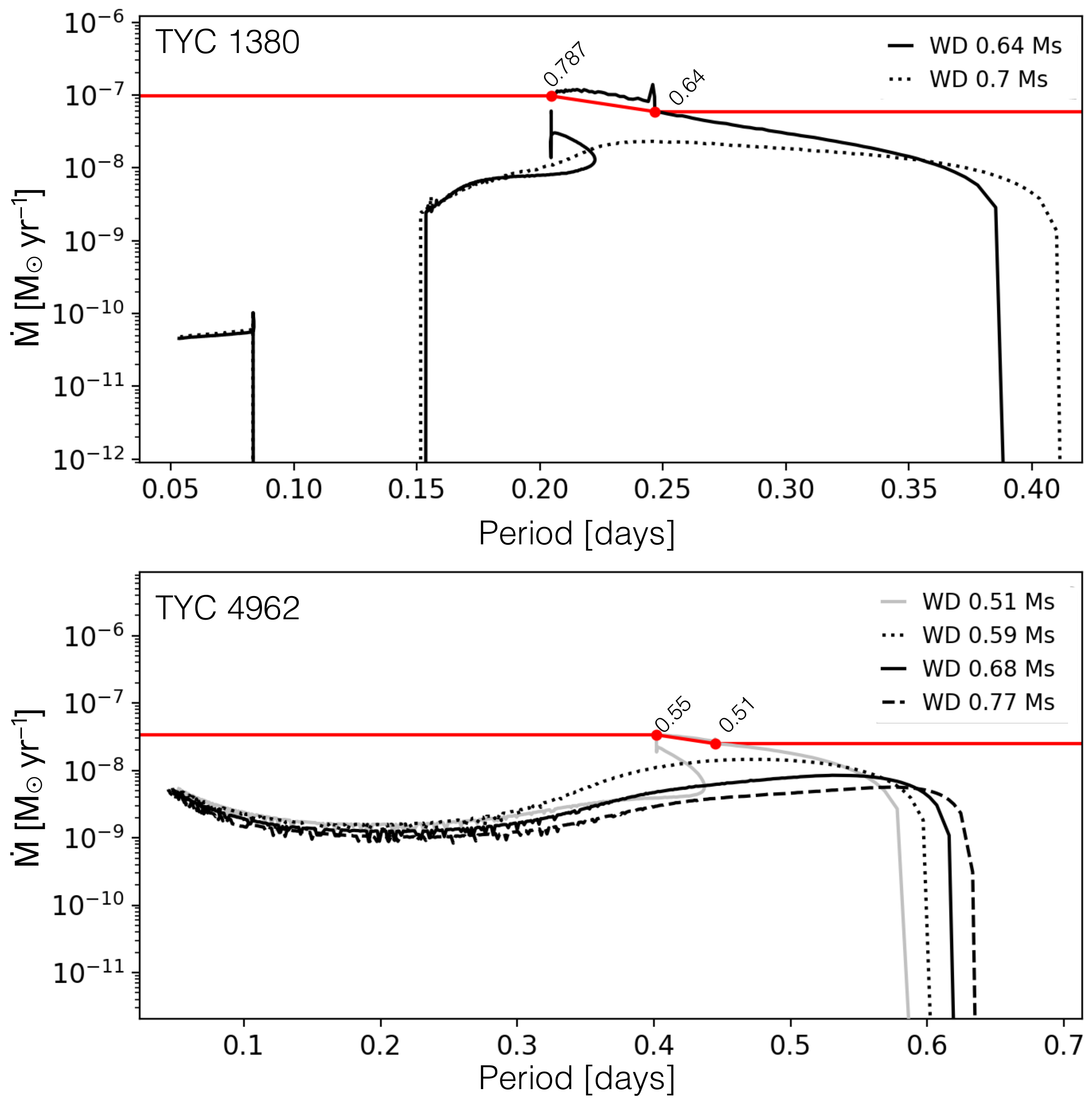}
    \caption{The future evolution of TYC\,1380 and TYC\,4962 for two and four different white dwarf masses respectively. While TYC\,1380 will evolve into a normal CV after perhaps passing through a period of high mass transfer (above the  red line) which would convert the system into a super soft source for $\simeq1.7$\,Myr. TYC\,4962 will become a CV with an evolved donor star and therefore, in contrast to CVs with unevolved donors, not detach in the orbital period range of $2-3$\,hr. This system will only appear as a super soft source if the white dwarf mass is smaller than the value estimated from $v\sin(i)$ (grey line, bottom panel). }
    \label{fig:mdot}
\end{figure}

PCEBs with more massive secondary stars are crucial for our understanding of close white dwarf binary formation. Their total mass exceeds the Chandrasekhar limit and therefore, in contrast to most white dwarf plus M-dwarf binaries, some of these systems might 
be potentially SN\,Ia progenitors.  
The future of these PCEBs with solar type secondary stars, however, is more diverse than that of PCEBs with late-type secondary stars, which virtually all become CVs sooner or later (in some cases it may take several Hubble times).
The future of PCEBs with G type secondary stars 
depends on the orbital period at the end of common envelope evolution,  
on the angular momentum loss mechanism that drives the system to shorter orbital periods, 
on the mass ratio of the binary and on the evolutionary status of the secondary star. The richness and variety of evolutionary futures of close WD+AFGK binaries is illustrated by the three systems characterized in this work. 
All three systems contain a G-type secondary star and a white dwarf with a mass typical for these objects  
and a short orbital period between $1.2$ and $2.5$~days. Despite these similarities, they face very different futures. 

In order to estimate the future of these systems in more detail, we performed dedicated simulations using \textsc{MESA} \citep{paxton11} and the same set-up and the same criteria for stable mass transfer as in \citet{parsons15}.  
For the current parameters we adopt the parameters listed 
in Table\,\ref{tab:stellarP}. The white dwarf mass estimates derived from the observations are quite uncertain. We therefore experimented with different white dwarf masses covering the range of masses and the minimum mass given in Table\,\ref{tab:stellarP}.
We discuss in more details simulations for different white dwarf masses which have been chosen to illustrate how the future evolution depends on this parameter. 

\subsubsection{TYC\,1380-957-1: potentially the second known super soft source progenitor}

Simulations for TYC\,1380 were performed assuming 
$M_{\mathrm{MS}}=1.18~\Msun$ and different white dwarf masses covering the range constrained by our observations, i.e. we used $0.59$, $0.64$, $0.65$, $0.68$, $0.7$, $0.73$, $0.85$~\Msun.
The current orbital period of the system 
is $P_\mathrm{orb}=1.6127$\,days. 
Two evolutionary tracks are shown in the upper panel of 
Fig.\,\ref{fig:mdot}. 
We find that the future of the system depends sensitively on the white dwarf mass. For the 
white dwarf masses exceeding $\simeq0.64$~\Msun the mass transfer remains dynamically and thermally stable and the system evolves into a CV. 
As an example for this evolution Figure\,\ref{fig:mdot} shows the track calculated with a white dwarf mass of $0.7$~\Msun. In this case, 
in $1.2$\,Gyr from now, when the orbital period has reduced to $9.84$\,hrs, mass transfer starts but never reaches values large enough to ignite stable nuclear burning on the surface of the white dwarf. Nova eruptions should therefore occur and the mass of the white dwarf remains nearly constant (non-conservative mass transfer). Thus, if the real white dwarf mass is $\gappr0.65$~\Msun, TYC\,1380 would simply be a pre-CV with a large secondary mass.  

However, if the white dwarf mass is $0.64$~\Msun\ 
 or smaller, the mass ratio becomes large enough to trigger thermal time scale mass transfer and stable hydrogen burning on the surface of the white dwarf.  This future seems, however, somewhat less likely because thermal time scale mass transfer will only occur when the white dwarf mass is at the lower limit of the mass range derived from the $v\sin(i)$ measurement. 
The corresponding evolutionary track is also shown in Fig.\,\ref{fig:mdot}. In $\sim$1.1\,Gyr from now the system will appear as a super soft source. 
During this phase we assume that 90 per cent of 
transferred mass remain on the white dwarf which therefore increases in mass. 
While the assumption of mass growth by 90 per cent of the accreted mass is arbitrary, we performed several numerical experiments with different mass growth fractions and found that the outcome of the evolution of this system does not sensitively depend on this parameter.  
When the white dwarf mass reaches $\sim0.79~\Msun$ at an orbital period of $\sim4.8$\,hrs, the mass ratio becomes small enough for stable angular momentum loss driven mass transfer and the system becomes a CV. 
Thus, while TYC\,1380 will in any case become a CV, it  might evolve through a short super soft X-ray binary phase if the
white dwarf mass is equal to or smaller than $\sim0.64$~\Msun. 

\subsubsection{TYC\,4700-815-1: unstable mass transfer and stellar merger} 

The future of TYC\,4700 is straight forward to predict. Given its current orbital period of $P_\mathrm{orb}=2.4667$\,days, the mass ratio of $3.3<q<3.8$ and the secondary mass and radius of $1.45~\Msun$ and $2.2~\Rsun$ 
the system will run into dynamically unstable mass transfer when the secondary fills its Roche-lobe in $490$~Myr from now when the orbital period will be $1.53$~days. This is predicted by our \textsc{MESA} simulations in agreement with \citet[][their figures 8 and 9]{geetal15-1}. 
The orbital energy 
of the binary will not be sufficient to expel the envelope and therefore the two stars will merge and evolve into a giant star with the core of this giant being the white dwarf we observe today. This giant will then evolve into a single white dwarf as the final fate of TYC\,4700.  

\subsubsection{TYC\,4962-1205-1: a progenitor of a CV with evolved donor} 

For TYC\,4962 we adopted  M$_{\mathrm{MS}}$\,=\,0.969~M${_\mathrm{\odot}}$ 
and three white dwarf masses corresponding to the range we estimated from the observations  and one more closer to the minimum mass, i.e. we assumed 
M$_\mathrm{{WD}}$\,=\,{\bf $0.51$, }$0.59$, $0.68$  and  $0.77$~M$_\mathrm{\odot}$.
The current orbital period of the system is P$_\mathrm{{orb}}$\,=\,$1.2798$\,days. The radius we estimated for the secondary star,  R$_\mathrm{MS}$\,=\,$1.4$~R$_\mathrm{\odot}$, implies that it has significantly evolved from the ZAMS. 
%which corresponds to the mean values of the ranges in agreement with %our estimations in \ref{tab:stellarP}.  
The starting model for our \textsc{MESA} simulations was obtained by evolving the secondary until it reaches the required radius 
of the evolved secondary star. 
According to this calculation the current age of the object is $6.5$~Gyr and the secondary star has a temperature of $6172$~$\mathrm{K}$.
Driven by angular momentum loss through magnetic braking the system will start mass transfer in $335$\,Myr at an orbital 
period of $\sim0.6$~days somewhat depending on the white dwarf mass. 
Evolving the system further we found that, 
independent of the mass of the white dwarf, the system evolves into a 
CV with a significantly evolved secondary star. 
In such systems the secondary does not become fully convective and therefore mass transfer is driven by magnetic braking until it reaches the orbital period minimum. The system does not evolve as detached binary through the orbital period gap
(see bottom panel of Fig\,\ref{fig:mdot}). 
%While this prediction appears entirely reasonable, 
%We note that 
This evolutionary path for CVs with evolved secondary stars has been predicted by \citet{Kolb+Baraffe00} and 
several such CVs have been found \citep[e.g.][]{thorstensenetal02-1,rebassa-mansergasetal14-1}. 
 As our white dwarf mass estimates based on $v\sin(i)$ is somewhat uncertain, we also ran a simulation assuming  a smaller white dwarf mass. If the white dwarf mass is as low as $0.51$M$_\mathrm{\odot}$, thermal time scale mass transfer will be triggered and stable hydrogen burning on the surface of the white dwarf will begin at a period of $0.44$\,days. This thermal time scale mass transfer will then continue for the next $582$\,Myr, until the white dwarf reaches the mass of $0.55$ M$_\mathrm{\odot}$. The system will then become a cataclysmic variable with an evolved donor secondary star. Given the small white dwarf mass that is required for this path to occur, we consider this scenario rather unlikely.  
For completeness, the corresponding simulation is nevertheless shown in Fig.\,\ref{fig:mdot}. 
%We did not use the minimum white dwarf mass calculated for TYC\,4962, primarily because theoretical models do not exist for stable hydrogen burning for low-mass white dwarfs (M$\mathrm{_{WD}}$ < 0.51 M$_\mathrm{\odot}$). Therefore, we decide to use the lowest white dwarf mass shown in \citet{Wolf13}. }
In any case, TYC\,4962 is the first known progenitor of a CV with evolved donor. The future evolution is driven by the nuclear evolution of the secondary star and therefore virtually independent on the white dwarf mass. 
%very similar for 
%different 

While it is clear that TYC\,4962 will evolve into a CV with an evolved donor star, we note a minor inconsistency in our calculations. The current temperature of the secondary is much smaller than that of an evolved single star with the same mass. To match the observed mass and radius in the MESA calculations, we had to accept a temperature that clearly exceeds the observed value. In close binaries such as the systems presented here, the secondary star's rotation is synchronised with the orbital motion. The resulting fast rotation of the secondary star 
offers an explanation for the discrepancy between the MESA models and the measured temperature. 
Fast rotating stars are potentially very active which implies that they are probably larger and cooler than a single stars of the same mass. Therefore, also the evolutionary time the system needed to reach the radius we derived from the observations, was probably significantly less than the $6.5$\,Gyr we obtained with MESA for a single star.

\subsection{Excluding contamination from active M-dwarf companions}

In \citet{parsons15} and \citetalias{parsons16} we presented \textit{HST} spectra of ten of our RAVE candidates with the aim of 
confirming that the 
UV excess is, at least in most cases, indicative of a white dwarf. We found that this is indeed the case: nine out of the ten \textit{HST} spectra confirmed the presence of a white dwarf and in one case the UV excess came from a hot subdwarf or pre-helium white dwarf. While this experiment clearly showed that our target selection works very well, 
for individual objects, we cannot per se exclude the possibility that an unseen flaring/active M-dwarf companion is producing the UV excess instead of a white dwarf. 

However, flaring M-dwarfs are very 
unlikely to produce the \textit{GALEX} fluxes of our three targets. Assuming that the optically unseen stellar component is an M-dwarf with $T_{\mathrm{eff}}=4000$~K and $\log g=4.6$, according to the BT-NextGen spectral library \citep{Allard12} the magnitude difference between the M\,dwarf model and the observed NUV magnitude is $9.51$, $11.7$ and $8.81$  
for TYC\,1380, TYC\,4700 and TYC\,4962, respectively. 
These values are far beyond the highest values of the observed $\Delta m_\mathrm{NUV}$ distribution found 
for flaring M-dwarfs \citep{welsh07}. 

\begin{figure}
    \centering
    \includegraphics[width=1\columnwidth]{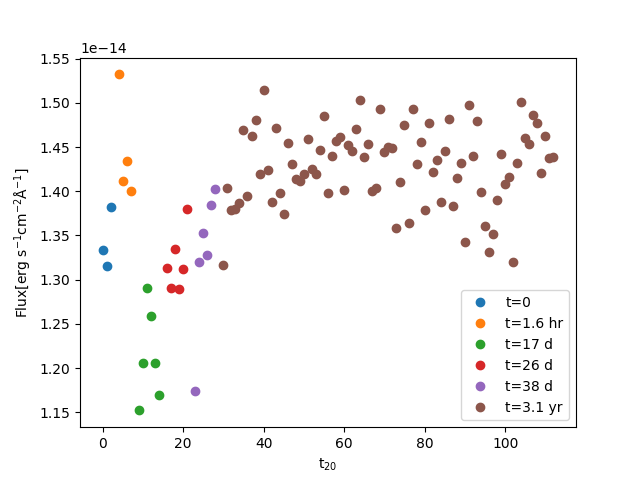}
  \caption{NUV light curve of TYC\,1380. Each color represent one of the six exposures. The value of $t$ represent the starting time of the exposure relative to the fist exposure marked in blue, while t$_{20}$ stands for $20$~second bins.}
    \label{fig:lc_tyc1380}
\end{figure}

For one of the systems discussed in this paper, we could complement the above finding with the \textit{GALEX} light curve. We used {\sc gPhoton} \citep{gphoton} to produce an NUV light curve for TYC\,1380 with $20$~second binning in each exposure and an aperture radius of $0.004^{\circ}$ and the resulting light curve is shown in Fig.\,\ref{fig:lc_tyc1380}. 
Taking into account that the fluxes represented by the green dots correspond to observations where the position of the system was close to the edge of the \textit{GALEX} detector and are therefore likely compromised, 
Fig.\,\ref{fig:lc_tyc1380} clearly excludes that the 
UV excess in TYC\,1380 is the result of a flaring M-dwarf.  

\begin{figure}
    \centering
    \includegraphics[width=1\columnwidth]{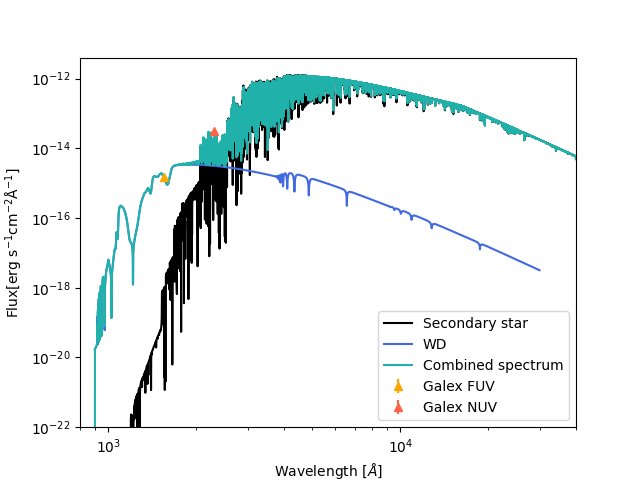}
  \caption{Best white dwarf model fit to the observed UV excess in TYC\,4962. The synthetic spectrum of the white dwarf from \citet{koester10-1} (shown in blue) corresponds to T$_{\mathrm{eff}}=11\,000$~K, $\log{g}=8$, and M$=0.6\,\mathrm{M}_{\odot}$. While this is clearly just one possible solution, it illustrates that the UV excess can be perfectly explained by the presence of a white dwarf. }
    \label{fig:spec_fit1}
\end{figure}

As a final test, we check whether a white dwarf model can produce the observed UV excess in TYC\,4962 which is the system with the smallest UV excess of the three systems we discussed in this work. 
Using a chi-square fitting and interpolating the UV magnitudes of the cooling models\footnote{http://www.astro.umontreal.ca/~bergeron/CoolingModels/} 
of \cite{holberg+bergeron06-1}, \cite{kowalski+saumon06-1} and \cite{tremblayetal11-2}, we found the 
best-fit stellar parameters and distance of the white dwarf.  
Figure \ref{fig:spec_fit1} shows that the UV excess is perfectly matched (p-value<$0.02$) using the white dwarf model spectrum with  \textbf{T$_{\mathrm{eff}}=11\,000$}~K, $\log{g}=8$, M$=0.6~\mathrm{M}_{\odot}$, at the distance measured with \textit{GAIA}. While this fit is highly degenerate -- we only have two data points in the UV -- it clearly shows that a white dwarf model can reproduce the UV excess. 

We therefore conclude that the presence of a white dwarf in the three targets presented here is by far the most reasonable interpretation of the UV excess.

\section{Conclusions}

We presented the identification of three close binaries containing a G-type star and a white dwarf. The presence of the latter is indicated by a UV excess detected with \textit{GALEX}. Based on measurements of the orbital and stellar parameters of these systems, we reconstructed their past and predicted their future. We find that all three systems have most likely formed through common envelope evolution. In contrast to previously known PCEBs with G-type companions, in the three systems presented here we do not find any indications for additional sources of energy acting during common envelope evolution. Despite having similar stellar masses and orbital periods, the future of all three systems is different. TYC\,4962 will evolve into a cataclysmic variable (CV) star with an evolved donor, TYC\,4700 will not survive the next phase of mass transfer and evolve into a single white dwarf,
and TYC\,1380 will perhaps appear as a super soft X-ray source before becoming a rather typical CV. This shows that the future evolution as well as the past of PCEBs with AFGK secondary stars is more complicated than those of PCEBs with M-dwarf companions. We therefore conclude that the identification and observational characterization of more PCEBs consisting of a white dwarf with an AFGK companion is likely to provide crucial constraints of white dwarf binary evolution theories with potential implications for our understanding of the progenitors of thermonuclear supernovae.

\section*{Acknowledgements}

MSH acknowledges support through a Fellowship for National PhD students from ANID, grant number 21170070. MRS acknowledges financial support from FONDECYT (grant 1181404).
SGP acknowledges the support of the STFC Ernest Rutherford Fellowship. 
BTG was supported by the UK STFC grant ST/T000406/1.
FL acknowledges support from the ESO studentship program.
RR has received funding from the postdoctoral fellowship program Beatriu de Pin\'os, funded by the Secretary of Universities and Research (Government of Catalonia) and by the Horizon 2020 program of research and innovation of the European Union under the Maria Sk\l{}odowska-Curie grant agreement No 801370. OT was supported by a Leverhulme Trust
Research Project Grant. GT acknowledges support from PAPIIT project IN110619.
MZ acknowledges support from CONICYT PAI (Concurso Nacional de Inserci{\'o}n en la Academia 2017, Folio 79170121) and CONICYT/FONDECYT (Programa de Iniciaci{\'o}n, Folio 11170559). ARM acknowledges support from the MINECO under the Ram\'on  y Cajal programme (RYC-2016-20254) and the AYA2017-86274-P grant, and the AGAUR grant SGR-661/2017. JJR acknowledges the support of the National Natural Science Foundation of China 11903048 and 11833006. We acknowledge the support of the staff of the Xinglong 2.16m telescope. This work was partially supported by the Open Project Program of the Key Laboratory of Optical Astronomy, National Astronomical Observatories, Chinese Academy of Sciences. The results presented in this paper are partly based on observations collected at the European Southern Observatory under program IDs  0102.D-0355(A) and 0101.D-0415(A).

\section*{Data Availability}
The data underlying this article will be shared on reasonable request to the corresponding author.

%%%%%%%%%%%%%%%%%%%%%%%%%%%%%%%%%%%%%%%%%%%%%%%%%%

%%%%%%%%%%%%%%%%%%%% REFERENCES %%%%%%%%%%%%%%%%%%

% The best way to enter references is to use BibTeX:

%\bibliographystyle{mnras}
%\bibliography{example} % if your bibtex file is called example.bib

\bibliographystyle{mnras}
\bibliography{TWBPS_IV_MSH}

%%%%%%%%%%%%%%%%%%%%%%%%%%%%%%%%%%%%%%%%%%%%%%%%%%

%%%%%%%%%%%%%%%%% APPENDICES %%%%%%%%%%%%%%%%%%%%%

\appendix
\section{Radial velocities measured for all three systems}
% Example table
\begin{table}
    \setlength{\tabcolsep}{4pt}
	\centering
	\caption{TYC 4700-815-1 radial velocity measurements.}
	\label{tab:TYC4700}
	\begin{tabular}{llllll} % four columns, alignment for each
		\hline
		Instrument  & Exptime & S/N & BJD & RV & error\\
		            &    [$s$]& & & [$Km~s^{-1}$] & [$Km~s^{-1}$]\\
		\hline
HRS  &900  & 40	 &2458092.06920 &-25.30 &  	0.49 \\
HRS  & 900  & 35	 &2458175.99240 &-29.67 &  	0.72 \\
HRS	 &430   & 28  &2458130.08178	      &-5.03 & 0.77\\
HRS	 & 560  & 40	 &2458169.96268      &28.34 &	0.50 \\
HRS	 & 560  & 40   &2458169.97201	      &30.34	& 0.48\\
HRS	 & 560  & 40   &2458169.98144       &30.07	& 0.65 \\
HRS	 & 560  & 40	 &2458169.99050	      &30.22	& 0.59 \\
HRS	 & 500  & 45	 &2458170.95383	      &-18.84 & 0.50 \\
HRS	 & 400 & 53	 &2458170.96179      &-21.31	& 0.68\\
HRS	 & 240  & 37	 &2458170.96850	      &-19.33 & 	0.68 \\
HRS	 & 200  & 37   &2458170.97399	      &-20.22 &	0.76 \\
HRS	 &  240 & 35   &2458171.94807      &-13.36 &	0.65\\
HRS	 & 480  & 35	 &2458171.95351	      &-12.85 &	0.70 \\
HRS	 & 480  & 25  &2458171.96850       &-9.26 &	1.22 \\
HRS	 & 540  & 35   &2458171.97273	      &-10.59 &	0.60 \\
HRS	 & 540  & 18	 &2458175.95720      &-24.88 &	2.96 \\
HRS	 &  240 & 30   &2458175.96460	      &-28.19 &	1.13 \\
HRS	 &  120 & 30   &2458175.97270       &\-29.59 &	2.14\\
HRS	 & 240  & 35 &2458175.98238      &-28.99 &	0.70 \\
MRES &  120 & 21	 &2458184.03678      &-30.60 &	1.81\\
MRES & 240 & 28   &2458184.04033        &-33.22 &	1.34 \\
MRES & 60   & 25 &2458185.01066	      &36.60	 &  1.87 \\
MRES &  60  & 20   &2458185.02971        &37.64	 &  2.27 \\
MRES  & 150 & 24  &2458185.04959        &35.81  &	1.61 \\
FEROS & 120 &39   &2458486.59061272  &-11.03	 & 0.17  \\
FEROS & 120 &37   &2458486.66071053  &-17.01	 & 0.20 \\
FEROS & 120 & 60  &2458486.66322596 &-18.54     &	0.20 \\
FEROS &120 & 48  &2458487.63613462 &-19.36	 &  0.14 \\
FEROS & 120 & 57  &2458487.68227370  &-14.85	 & 0.14  \\
FEROS & 120 & 53  &2458488.61826496  &27.69      &0.15 \\
FEROS & 120 & 38  &2458488.67449235  &24.83     &  0.13 \\
FEROS & 120 & 23   &2458489.60644339  &-45.69     &	0.20 \\
FEROS & 120 & 53   &2458489.67938950  &-44.66	 & 0.27 \\
FEROS &120 & 42	 &2458491.64877842  &-22.43	 & 0.15 \\
FEROS & 120 &34    &2458491.67602759  &-26.08     &	0.17 \\
FEROS & 120 & 41	 &2458491.69341871  &-26.29     &	0.17 \\
FEROS & 120 & 40	 &2458492.65260996  &-10.47     &	0.19 \\
FEROS & 120 & 36   &2458492.67067858  &-9.11	     & 0.18 \\
FEROS & 120 & 30	 &2458492.68748925  &-8.27	 & 0.17 \\
UVES  &  60 & 93   &2458353.845644055 &-40.01	& 1.36 \\
UVES  & 60  & 37    &2458390.870005791 &-43.15 & 1.96 \\
ESPRESSO & 1200 & 20 &2458802.8106283 &-37.05  &	2.62 \\
ESPRESSO & 1200 & 19 &2458803.8033296 &23.41 & 2.39  \\

		\hline
	\end{tabular}
\end{table}
\begin{table}
    \setlength{\tabcolsep}{4pt}
	\centering
	\caption{TYC 1380-957-1 radial velocity measurements.}
	\label{tab:TYC1380}
	\begin{tabular}{llllll} % four columns, alignment for each
		\hline
		Instrument  & Exptime& S/N &BJD& RV & error\\
		            &  [$s$] & & & [$Km~s^{-1}$] & [$Km~s^{-1}$]\\
		\hline
HRS	&  1200 & 35  &2458171.10569 & 49.36 &	1.0 \\
HRS	& 1400 & 35  &2458176.20292& -17.56 &1.2 \\
HRS	&  2000 & 26  &2458243.03968 & -4.99	&2.2 \\
ESPRESSO &	1200 &18 &2458390.9747601 &	-69.82	&   1.19 \\
ESPRESSO &	1200 &19 &2458391.0227932 &	-72.73	& 0.92 \\
ESPRESSO &	1200 &21 &2458391.9867707 &	62.85	& 1.02 \\
ESPRESSO &  1200 &22 &2458450.9844952 &	-21.61 & 1.34 \\
ESPRESSO &  1200 &20 &2458451.9935934 &	-18.70	& 2.16 \\
FEROS	& 120 &18	&2458486.68750550 &	32.12&	0.13\\
FEROS	& 120 &16	&2458486.70998961 &	37.63	&0.13 \\
FEROS	& 120 &17	&2458487.66240839 &	-62.96 &	0.10 \\
FEROS	&120 &16	&2458487.75553172 &	-71.04 &	0.10 \\
FEROS	& 120&31	&2458488.66635571 &	72.53	 &0.10  \\
FEROS	& 120 &21 &2458489.66670363 &	-35.87 &	0.17 \\
FEROS	& 240 &29	&2458490.76647412 &	-39.85 &	0.09 \\
FEROS	& 240 &30 &2458490.77041791 &	-40.92 &	0.09 \\
FEROS	& 240 &10	&2458490.77430824 &	-42.00 &	0.09 \\
FEROS	& 240 &25	&2458490.77819312 &	-42.96 &	0.09 \\
FEROS	& 240 &22 &2458490.78256429 &	-43.78 &	0.09 \\
FEROS	& 240 &25	&2458490.78825937 &	-44.96 &	0.09 \\
FEROS	& 120 &23	&2458490.83072595 &	-53.54 &	0.13 \\
FEROS	&120 &24	&2458491.67514523 &	63.23 &	0.11 \\
FEROS	& 120 &24	&2458491.68357584 &	64.50	& 0.11 \\ 
FEROS	& 120 &12 &2458491.77918888 &	73.61	& 0.14 \\
FEROS	& 120 &16	&2458492.66294326 &	-70.92 &	 0.11 \\
FEROS	& 120 &17	&2458492.69528291 &	-69.29 &	0.10  \\
FEROS	& 120 &19	&2458492.73411412 &	-65.38 &	0.09 \\
FEROS	& 120 &13	&2458492.75539783 &	-62.74 &	0.12 \\
FEROS	& 120 &13	& 2458492.78775605 &	-58.20 &	0.08 \\
FEROS	& 180 &15	&2458600.50883313 &	 -55.86 &	0.10 \\
FEROS	& 180 &21	&2458601.49588304 &	75.18	& 0.09  \\
FEROS	& 180 &20	&2458602.49497791 &	-49.29 &	 0.07 \\
FEROS	& 180 &22	&2458603.53605945 &	-8.12	& 0.09 \\
FEROS	& 180 &22	&2458604.49805593 &	51.15	&0.12  \\
FEROS	& 180 &19	&2458605.48214311 &	-70.92 & 0.09 \\
FEROS	& 240 &24	&2458606.48415264 &	61.11	& 0.08 \\
FEROS	& 240 &12	&2458606.48754113 &	60.38	& 0.07 \\

		\hline
	\end{tabular}
\end{table}

\begin{table}
    \setlength{\tabcolsep}{4pt}
	\centering
	\caption{TYC 4962-1205-1 radial velocity measurements.}
	\label{tab:TYC4962}
	\begin{tabular}{llllll} % four columns, alignment for each
		\hline
		Instrument & Exptime &S/N & BJD & RV & error\\
		            &    [$s$]&& & [$Km~s^{-1}$] & [$Km~s^{-1}$]\\
		\hline

FEROS		& 120 &49   &2458290.63890684  &	10.55 & 	0.04 \\
FEROS		& 120 &54   &2458290.64296182  & 	9.74&	 0.04 \\
FEROS		& 120 &51   &2458290.51895498  &	26.46 &	0.04\\
FEROS		& 300 &69   &2458290.61714122  &	14.96 	&0.04\\
FEROS		& 300 &61  &2458290.62804077  &	12.82 &	0.04 \\
FEROS		& 120 &61   &2458290.63212027  &	11.97 &	0.06\\
FEROS		& 180 &65   &2458290.47164068  &	27.18 &	0.13\\
FEROS		& 180 &64   &2458292.54191363 &	-88.89 &	0.06\\
FEROS		& 180 &66   &2458292.54469778 &	-87.98 &	0.05\\
FEROS		& 180 &67   &2458292.64038933 &	-65.31 &0.06\\
FEROS		& 180 &65   &2458292.64444679 &	-64.99 &	0.05\\
FEROS		& 300 &6   &2458293.48043724 &	-70.12 &	0.06\\
FEROS		& 300 &56   &2458293.48450716 &	-71.74 &	0.31\\
FEROS		& 300 &39   &2458293.64986508 &	-99.76 & 0.07\\
FEROS		& 300 &55   &2458293.65394596 &	-99.83 &	0.07\\
FEROS		& 300 &95   &2458294.47693482 &	11.14 & 0.06\\
FEROS		& 120 &61   &2458294.57760653 &	-18.89 &	0.07\\
FEROS		& 360 &53   &2458294.65813593 &	-37.52 &	0.05\\
FEROS		& 120 &49   &2458295.56410647  &	24.82 &	0.05\\
FEROS		& 300 &59   &2458295.62289638 &	26.42 &	0.05\\
FEROS		& 300 &80  &2458600.53527200 &	-40.52 &	0.05\\
FEROS		& 120 &74   &2458600.61655423 &	-69.08 &	0.05\\
FEROS		& 300 &68  &2458601.52255787 &	24.66	& 0.06\\
FEROS		& 300 &48   &2458601.64901651 &	4.00 &	0.06\\
FEROS		& 300 &67   &2458602.77673050 &	26.15 &	0.05\\
FEROS		& 120 &68  &2458603.55421825 &	-81.95 &	0.05\\
FEROS		& 300 &69   &2458604.61991781 &	-99.22 &	0.07\\
FEROS		& 300 &65   &2458604.65812713 &	-101.21 &	0.07\\
FEROS		& 300 &62  &2458606.51420823 &	 22.62 &	0.05 \\
FEROS		& 120 &59   &2458606.51687580 &	 22.95 & 0.05 \\
FEROS		& 120 &64   &2458606.51954985 &	 23.15 &	0.05 \\
FEROS		& 120 &70   &2458606.52224277&	 23.40 & 0.05 \\
FEROS		& 120 &68   &2458606.52493824 &	 23.68 &	0.05 \\
ESPRESSO	& 1200 &30	&2458857.0716883 &	-43.47 &0.91\\
ESPRESSO	& 1200 &32	&2458860.0296351 &	26.28	& 1.19\\
ESPRESSO	& 1200 &28	&2458862.0141624 &	-93.56	&1.98\\
ESPRESSO	& 1200 &40	&2458915.0544167 &	27.21	&1.18\\
ESPRESSO	& 1200 &43	&2458916.9562919 &	-98.02	&1.56\\
ESPRESSO	& 1200 &39   &2458917.8290392 &	-17.92	&1.95\\

		\hline
	\end{tabular}
\end{table}

%%%%%%%%%%%%%%%%%%%%%%%%%%%%%%%%%%%%%%%%%%%%%%%%%%

% Don't change these lines
\bsp	% typesetting comment
\label{lastpage}
\end{document}